\documentclass[twocolumn,superscriptaddress,letter]{revtex4-2}%
\usepackage{eurosym}
\usepackage{amsbsy}
\usepackage{latexsym,epsfig}
\usepackage{graphicx}
\usepackage{dcolumn}
\usepackage{subfigure}
\usepackage{comment}
\usepackage{multirow}
\usepackage{color}
\usepackage{bm}
\usepackage{mathrsfs}
\usepackage{amssymb}
\usepackage{amsfonts}
\usepackage{amsmath}
\usepackage{xspace}
\usepackage{soul}
\usepackage{cancel}
\usepackage{float}
\usepackage{epstopdf}
\usepackage{tabularx}
\usepackage{longtable}
\usepackage[colorlinks=true, letterpaper=true, pdfstartview=FitV, linkcolor=red, citecolor=blue, urlcolor=blue]{hyperref}
\usepackage[normalem]{ulem}
\usepackage[version=4]{mhchem}
\newcommand{\be}{\begin{eqnarray}}
	\newcommand{\ee}{\end{eqnarray}}

\begin{document}
\title{Engineering quantum Mpemba effect by Liouvillian skin effect}
\author{Xiang Zhang}
\author{Chen Sun}

\author{Fuxiang Li}

\email[Corresponding author: ]{fuxiangli@hnu.edu.cn}
\affiliation{School of Physics and Electronics, Hunan University, Changsha 410082, China}
\date{\today}

\begin{abstract}
We propose a novel approach to engineer the quantum Mpemba effect (QME)---wherein an initial state farther from the steaty state relaxes faster than a closer one---by the Liouvillian skin effect (LSE) in open quantum systems. We show that, in the open quantum chain with LSE, QME can be easily realized by considering only the spatial profile of initial states, since the initial states localized on the left or right edges experience distinctive relaxation process (algebraic or exponential decay). This approach circumvents
the necessity of careful initial-state design and fine-tuning
of control parameter.
% This choice doesn't have to consider how to eliminate the slowest damping mode, as was done in unitary transformation.
%Focusing on the quadratic Lindbladians, we consider two concrete cases to design the initial states, thereby realizing the QME. 
Moreover, when the initial correlation matrix contains off-diagonal elements, we uncover a new kind of QME  which manifest as two crossings in the Hilbert-Schmidt distance at different times.
This work unveils the deep connection between QME and LSE, and  provides a physically intuitive understanding of QME and straightforward pathway for the initial state preparation thereby enabling readily accessible experimental preparation.
%The connection between the LSE and QME provides a physically more intuitive understanding of the QME. Moreover, the LSE serves as an ideal platform for realizing the QME and the spatial profile of the LSE provides a straightforward pathway for the initial state preparation, thereby enabling readily accessible experimental preparation. 
\end{abstract}
\maketitle

{\it Introduction.--} The relaxation process in open quantum systems holds fundamental significance across both nonequilibrium physics and applied quantum technologies. Effectively steering the relaxation process could accelerate quantum computation \cite{degen2017quantumsensing, giovannetti2004, pirandola2018advances}, enable the development of faster search algorithms \cite{verstraete2009dissipation, cirac2012goals, lloyd1996universal}, and facilitate the preparation of quantum states essential for quantum-enhanced metrology \cite{wei2016duality, wild2016adiabatic}.
Among the various strategies for accelerating relaxation toward thermal equilibrium, a particularly intriguing approach is inspired by the so-called Mpemba effect, a counterintuitive relaxation anomaly in which a hotter system can cool faster than a colder one \cite{Mpemba1969cool,Kell1969freezing}.
This phenomenon was first understood in the context of classical nonequilibrium physics \cite{Lu2017nonequilibrium,Lasanta2017when1,Klich2019mpemba,Bechhoefer2021fresh,Van2025the,TEZA20261,Summer2025resource,Teza2023relax,Pemart2024short,Lapolla2020faster}, and then extended to quantum systems, in both open system \cite{Chatterjee2023quantum,Nava2024mpemba,Nava2025pontus,Moroder2024thermodynamics,Medina2025anomalous,Kochsiek2022acc,Ares2025quantum1, Strachan2025non,teza2025speedups,Bao2025acc,Solanki2025universal,Liu2025general,Chattopadhyay2026anomaly,Carollo2021exp,Das2025role,Qian2025intri,Nava2025speed,Bagui2025accelerated,Bagui2025detection,Ma2025quantum,Tejero2025asy,Zhang2025quantum,Alyuruk2025thermodynamic,Longhi2025quantum1} and closed systems \cite{Ares2023entanglement,Liu2024symmetry,Yu2025quantum,Turkeshi2025quantum,Saliba2025unraveling,Liu2025symmetry,Banerjee2025entanglement,Sugimoto2025prethermal,Rylands2024micro,Yamashika2024entang,Chatterjee2024mult,Yu2025tun}. When searching for the optimal initial state to accelerate relaxation, however,  the existing  implementations based on quantum Mempba effect (QME) typically require careful initial-state design and fine-tuning of control parameters \cite{Carollo2021exp,Das2025role,Qian2025intri,Nava2025speed,Bagui2025accelerated,Bagui2025detection,Ma2025quantum,Ares2025quantum, Yu2025tun}, or alternatively, rely on  temporarily coupling the system to a reset channel \cite{Bao2025acc,Solanki2025universal,Liu2025general}. 

As a seemingly unrelated phenomenon, the non-Hermitian skin effect refers to  the exponential localization of a vast majority of eigenstates near the boundaries of non-Hermitian systems \cite{Yao2018edge,Yao2018non,Song2019non1}. The non-Hermitian skin effect fundamentally alters the conventional bulk-boundary correspondence, and yields novel phenomena in wavefunction dynamics
such as unidirectional amplification and quantum sensing enhancement \cite{Kunst2018bio,Duo2021exa,Borgnia2020non,Zhang2020cor,Lee2019hyb,Li2020top,Yang2020non,Yi2020non,Okuma2020top,Shimomura2024gen,Longhi2025err,Zheng2024exact}. More recently, the skin effect was identified to exist in the Liouvillian superoperator due to its
intrinsic non-Hermiticity and was dubbed the Liouvillian skin effect (LSE) \cite{Haga2021lio,Begg2024qua,Mao2024lio,Yang2022lio,Hamanaka2023int,Wang2023acc,Cai2025optical}. It has also been pointed out that the LSE has a striking influence on the relaxation processes \cite{Song2019non,Haga2021lio,Mori2020resolving,McDonald2022non,Lee2023ano,Belyansky2025phase}. %However, the relation between LSE and QME has not been explored. 

% In this Letter, by uncovering the deep connection between QME and Liouvillian skin effect (LSE), we propose a straightforward and intuitive approach for selecting the initial states and realizing QME in open quantum systems.
In this Letter, we explore the spatial structure of relaxation process in open quantum system and illustrate that LSE enables a straightforward and intuitive approach for selecting the initial
states and realizes QME in open quantum systems. The initial state can be prepared simply by exploiting
the spatial distribution of the particle numbers induced by the LSE. This approach circumvents the necessity of careful initial-state design and fine-tuning of control parameter.
%{\color{red} Specifically, ...how to choose initial state??  } 
Moreover, when the initial state contains inter-site correlations, a new kind of QME may occur which consists of two crossings in the Hilbert-Schmidt distance.

% Leads to dramatic difference between OBC and PBC spectra and profoundly altering topological phenomena such as bulk-boundary corresponddence. Also yield novel phenomena in wavefunction dynamics such as unidirectional applification, and quantum sensing enhancement. On the other hand, non-Hermiticity naturally arises in the Lindblad master equation that governs the time evolution of density matrix under the Markovian approximation \cite{Lindblad1976generators,Gorini1976completely,Breuer2002theory}. It has been shown that the non-Hermitian Hamiltonian exhibits unique features, among which the non-Hermitian skin effect that describes the accumulation of eigenstates near the boundaries \cite{Yao2018edge,Yao2018non,Song2019non1} has attracted growing attention both theoretically \cite{Kunst2018bio,Duo2021exa,Borgnia2020non,Zhang2020cor,Lee2019hyb,Li2020top,Yang2020non,Yi2020non,Okuma2020top,Shimomura2024gen,Longhi2025err,Zheng2024exact} and experimentally \cite{Helbig2020generalized,Xiao2020non,Ghatak2020observation,Zou2021observation,Zhang2021observation,Liang2022dynamic}. {\color{red} Non-Hermitian skin effect also induces more intriguing phenomenon, including entanglement phase transitions \cite{Kawabata2023enta,Liu2024dynamical} and has been generalized to many body systems \cite{Shimomura2024general,Hu2025many,Gliozzi2024many,Lee2020many,Alsallom2022fate}.} Notably, a natural question is whether this influence on relaxation can be applied to the quantum Mpemba effect?

% {\it Markovian open quantum dynamics.--}
We shall start by reviewing the Lindblad master equation \cite{Lindblad1976generators,Gorini1976completely,Breuer2002theory} that describes the time evolution of the reduced density matrix $\rho$ of a  Markovian open quantum chain with size $L$: 
\begin{equation}
    \frac{d\rho}{dt}=-i[H,\rho]+\sum_{\mu}(2L_{\mu}\rho L_{\mu}^{\dagger}-\{L_{\mu}^{\dagger}L_{\mu},\rho\}),
    \label{lindblad}
\end{equation}
in which $H$ is the Hamiltonian  describing the unitary evolution, and $L_\mu$’s are the jump operators describing the incoherent dynamics induced by the coupling to the environment. The right-hand side of Eq.~(\ref{lindblad}) defines the Liouvillian superoperator $\hat{\mathcal{L}}$ (also called Lindbladian), acting on a doubled Hilbert space, and the master equation is conveniently written as $\partial_t|\rho\rangle\rangle=\hat{\mathcal{L}}|\rho\rangle\rangle$, where $|\rho\rangle\rangle$ now refers to the vectorized version of the reduced density matrix of dimension $2^L\times 2^L$ by flattening $|v_1\rangle\langle v_2|$ into $|v_1\rangle\otimes|v_2\rangle$ and the inner product in this linear space is defined as $\langle\langle A|B\rangle\rangle={\rm tr}(A^\dagger B)$. 

% $\mathcal{L}$ is always guaranteed to have a
% zero eigenvalue with the steady-state density matrix corresponding to the right eigenvector $|\rho_{ss}\rangle\rangle$.

\begin{figure}[htbp]
	\includegraphics[width=1\linewidth]{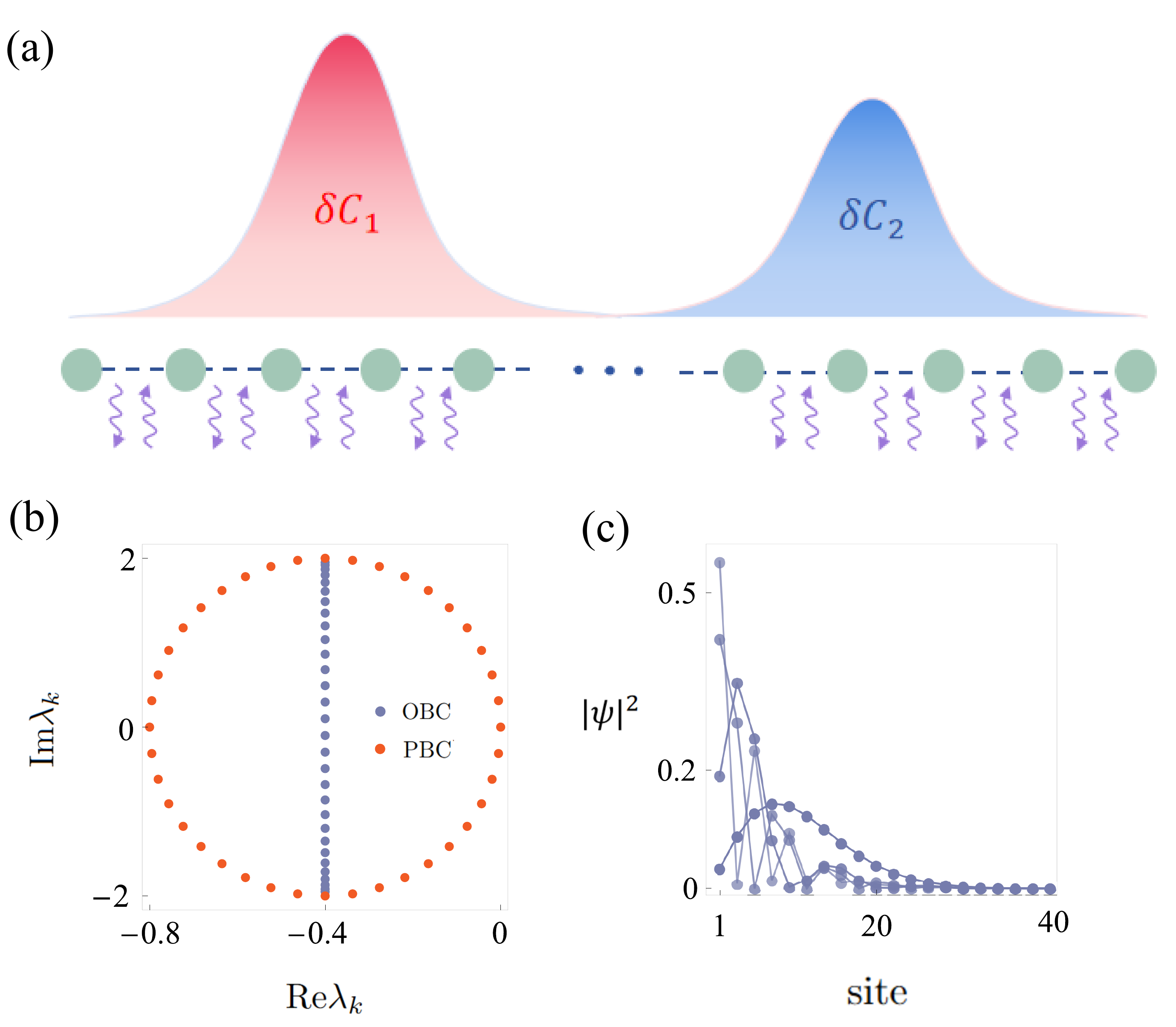}
	\centering
	\caption{Schematic illustration of the quantum Mpemba effect engineered by the Liouvillian skin effect. (a) Sketch of the model and the selection of the initial states $\delta C_1$ and $\delta C_2$ according to the spatial profile of localized eigenstates. 
    The purple arrows represent the linear gain and loss operators at each site. %(b) and (c) are the two typical QME (QME-I and QME-II). For (b), QME-I is realized when one unitary transforms the initial state $\rho_2$ to another state $\rho_1$ which makes the coefficient of the slowest mode disappear so that $\rho_1(t)-\rho_{ss}\propto {\rm exp}(t{\rm Re}\epsilon_3)$. For (c), QME-II can occur at early times even though asymptotic decay rates remain unchanged if $|a_2^c|>|a_2^f|$, where the coefficients $a_2^{c(f)}$ denote the overlaps between the initial state and the slowest modes for close (farther) state. 
    % (b) and (c) illustrate that QME can be realized by choosing the two initial states $\delta C_{1, 2}(0)$ selected according to the LSE. 
    % For comparison, QME-II can not appear (dashed lines in (e)) when the eigenstates are extended (dashed line in (d)).  
    (b) The eigenvalues  of the effective Hamiltonian $\mathcal{H}_{{\rm eff}}$ (\ref{xh}) for open boundary conditions and periodic boundary conditions.  (b) shows the Liouvillian skin effect, i.e., localization of eigenstates under OBC. Parameters are $J=1$, $\gamma_g=\gamma_l=0.2$ and $L=40$.} 
	\label{model}	
\end{figure} 

Formally, we can write down the general evolution of  density matrix $\rho(t)$ as
\begin{equation}   |\rho(t)\rangle\rangle=e^{\hat{\mathcal{L}}t}|\rho_0\rangle\rangle=|\rho_{ss}\rangle\rangle+\sum_{i=2}^{2^{2L}}e^{\epsilon_i t}\langle\langle l_i|\rho_0\rangle\rangle |r_i\rangle\rangle, 	\label{rhot}
\end{equation}
where $|\rho_0\rangle\rangle$ is the initial state, and $|\rho_{ss}\rangle\rangle$ is the steady state that is the right eigenoperator of the Liouvillian superoperator $\hat{\mathcal{L}}$ associated with the zero eigenvalue. $|r_i\rangle\rangle$ and $|l_i\rangle\rangle$ are the  right  and left eigenvectors corresponding to the eigenvalue $\lambda_i$ which satisfy $\hat{\mathcal{L}}|r_i\rangle\rangle=\epsilon_i|r_i\rangle\rangle$ and $\hat{\mathcal{L}}^\dagger|l_i\rangle\rangle=\epsilon_i^*|l_i\rangle\rangle$. The right and left eigenmodes  satisfy $\langle\langle l_i|r_j\rangle\rangle=\delta_{i, j}$. Suppose that all eigenvalues are arranged in descending order of their real parts: $0={\rm Re}\epsilon_1>{\rm Re}\epsilon_2\geq{\rm Re}\epsilon_3\geq...$, the spectral gap is $|{\rm Re}\epsilon_2|$ and defines the longest timescale in the system such that $\rho(t)-\rho_{ss}\propto {\rm exp}(t{\rm Re}\epsilon_2)$. 

In Ref.~\cite{Carollo2021exp}, the authors show that an exponential speed up to the steady state can be achieved by a unitary transform  of the initial state to another state whose coefficient of the slowest mode disappears so that $\rho(t)-\rho_{ss}\propto {\rm exp}(t{\rm Re}\epsilon_3)$, dubbed as strong QME. %Later, Ref.~\cite{Moroder2024the} showed that this method to speed up is also valid when the lowest eigenmodes of the Lindbladian form a complex conjugate pair.  Whereas this unitary transformation can be exactly constructed, it is, in practice, challenging to implement. In addition, 
QME can also occur at early times even though asymptotic decay rates remain unchanged if $|a_2^c|>|a_2^f|$, where the coefficients $a_2^{c(f)}$ denote the overlaps between the initial state and the slowest modes for closer (farther) states \cite{Longhi2025quantum}, which has an analogy in classical Mpemba effect \cite{Lu2017nonequilibrium}.

{\it Setup.--} 
Without loss of generality, we consider the quadratic  Hamiltonian $\mathcal{H}=\sum_{ij}c_i^\dagger H_{ij}c_j$ in terms of the femionic creation (annihilation) operators $c_i^\dagger$ ($c_i$). The jump operators are denoted as a linear single particle loss and gain, with loss operators $L_\mu^l=\sum_iD_{\mu i}^lc_i$ and gain operators $L_\mu^g=\sum_iD_{\mu i}^gc_i^\dagger$, where $D_{\mu i}$ are the complex coefficients \cite{Prosen2008third,Prosen2010spectral,Van2019dyna,McDonald2023third,Bardyn2013topology,Song2019non,Mao2024symmetry}.
 Note that similar discussions can be generalized to bosonic systems with linear jump operators (see SM for examples).

% With the above-mentioned settings, the steady state is Gaussian and completely determined by the steady-state correlation matrix: $C_{ss}^{i,j}=\langle c_i^\dagger c_j\rangle$. 
With linear jump operators, the Lindblad master equation is quadratic and  the evolution of density matrix can be captured by a non-Hermitian single-particle correlation matrix defined as $C_{ij}(t)=\mathrm{Tr}[c_i^\dagger c_j\rho]$. The time evolution of $C(t)$ reads \cite{Chaduteau2026lind,Lieu2020tenfold}:
\begin{equation}
    \frac{dC(t)}{dt}=i[H^T,C(t)]-\{ M_l^T+M_g,C(t) \}+2M_g,
    \label{ct}
\end{equation}
where $(M_g)_{ij}=\sum_\mu D_{\mu i}^{g*}D_{\mu j}^{g}$ and $(M_l)_{ij}=\sum_\mu D_{\mu i}^{l*}D_{\mu j}^{l}$. %Note that $M_l$ and $M_g$ are Hermitian matrices. 
Define the effective non-Hermitian Hamiltonian:
\be
    \mathcal{H}_{{\rm eff}}=iH^T-(M_l^T+M_g),
    \label{xf}
\ee
Eq.(\ref{ct}) can be recasted into 
\begin{equation}
    \frac{dC(t)}{dt}=\mathcal{H}_{{\rm eff}}C(t)+C(t)\mathcal{H}_{{\rm eff}}^\dagger+2M_g.
\end{equation}
The steady state correlation matrix $C_{ss}$, to which long time evolution of any initial state converges, is determined by $dC_{ss}/dt=0$, or the Lyaponov equation $\mathcal{H}_{{\rm eff}}C_{ss}+C_{ss}\mathcal{H}_{{\rm eff}}^\dagger+2M_g=0$. 
% Then the steady state $C_{ss}$ can be written as:
%\begin{equation}
 %   C_{ss}=\sum_{i,j}-\frac{|R^i\rangle\langle L^i|2M_g|L^j\rangle\langle R^j|}{\lambda_i+\lambda_j^*}.
%    \label{ss}
%\end{equation}
After obtaining the steady state $C_{ss}$, we introduce the deviation of correlation matrix as 
\begin{equation}
    \delta C(t)=C(t)-C_{ss},
\end{equation}
whose  evolution can be formally written as $ \delta C(t)=e^{\mathcal{H}_{{\rm eff}}t}\delta C(0) e^{\mathcal{H}_{{\rm eff}}^\dagger t}$, and further reads:
\begin{equation}
    \delta C(t)=\sum_{i,j}e^{(\lambda_i+\lambda_j^*)t}|R^i\rangle\langle L^i|\delta C(0)|L^j\rangle\langle R^j|.
    \label{dcf}
\end{equation}
 Here, $\lambda_i$ is the $i$-th eigenvalues and $|R^i\rangle$ ($|L^i\rangle$) the corresponding right (left) eigenvectors of effective hailtonian ${\cal H}_{\rm eff}$.
If there exists LSE, then $|R^i\rangle$ and $|L^i\rangle$ would be exponentially localized on the opposite ends of the chain.  From Eq.~(\ref{dcf}) one can expect to easily achieve QME  by choosing different spatial configurations for initial $\delta C (0)$ such that  the coefficient $\langle L^i|\delta C_0|L^j\rangle$ may be  drastically different. This idea shares the same spirit as the underling physics of QME \cite{Lu2017nonequilibrium}. However, we want to stress two points that make our discussions qualitatively different from QME-II. First, all the eigenvalues $\lambda_i$ of ${\cal H}_{\rm eff}$ may have the same real part, see Fig.~\ref{model} (b), which may render the argument based on the slowest decay mode in Eq.~(\ref{rhot}) inapplicable. Second, the system may be thermodynamically large ($L\rightarrow \infty$), therefore more complex relaxation processes may occur, and algebraic decay rather than exponential decay may play a decisive role in QME. 
%Since all expectation values follow from $C$ via Wick’s theorem, we focus on $\mathcal{H}_{{\rm eff}}$ rather than $\mathcal{L}$ in the following and the distance from the steady state is calculated directly in terms of the correlation matrix $C$.
%From Eq.~(\ref{dcf}), the relaxation behavior of open quantum system should be dominated by the real part of the eigenvalues. However, if the system possesses skin effect, then there emerges extra relaxation channel. In particular, if the right eigenstates $R^i$ are localized on the left side of the one-dimensional system, then the $L^i$ would be localized on the right end. Therefore, one can easily tune the relaxation process by choosing different initial conditions for $\delta C_0$ which lead to drastically different values of the term $\langle L^i|\delta C_0|L^j\rangle$, and thus achieve the goals of quantum Mpemba effect.  %Since all expectation values follow from $C$ via Wick’s theorem, we focus on $\mathcal{H}_{{\rm eff}}$ rather than $\mathcal{L}$ in the following and the distance from the steady state is calculated directly in terms of the correlation matrix $C$.

% \begin{figure}[htbp]
% 	\includegraphics[width=1\linewidth]{energy.pdf}
% 	\centering
% 	\caption{Eigenenergy for fermionic and bosonic systems. $N=40$.} 
% 	\label{en}	
% \end{figure} 

\begin{figure}[thbp]
\includegraphics[width=1\linewidth]{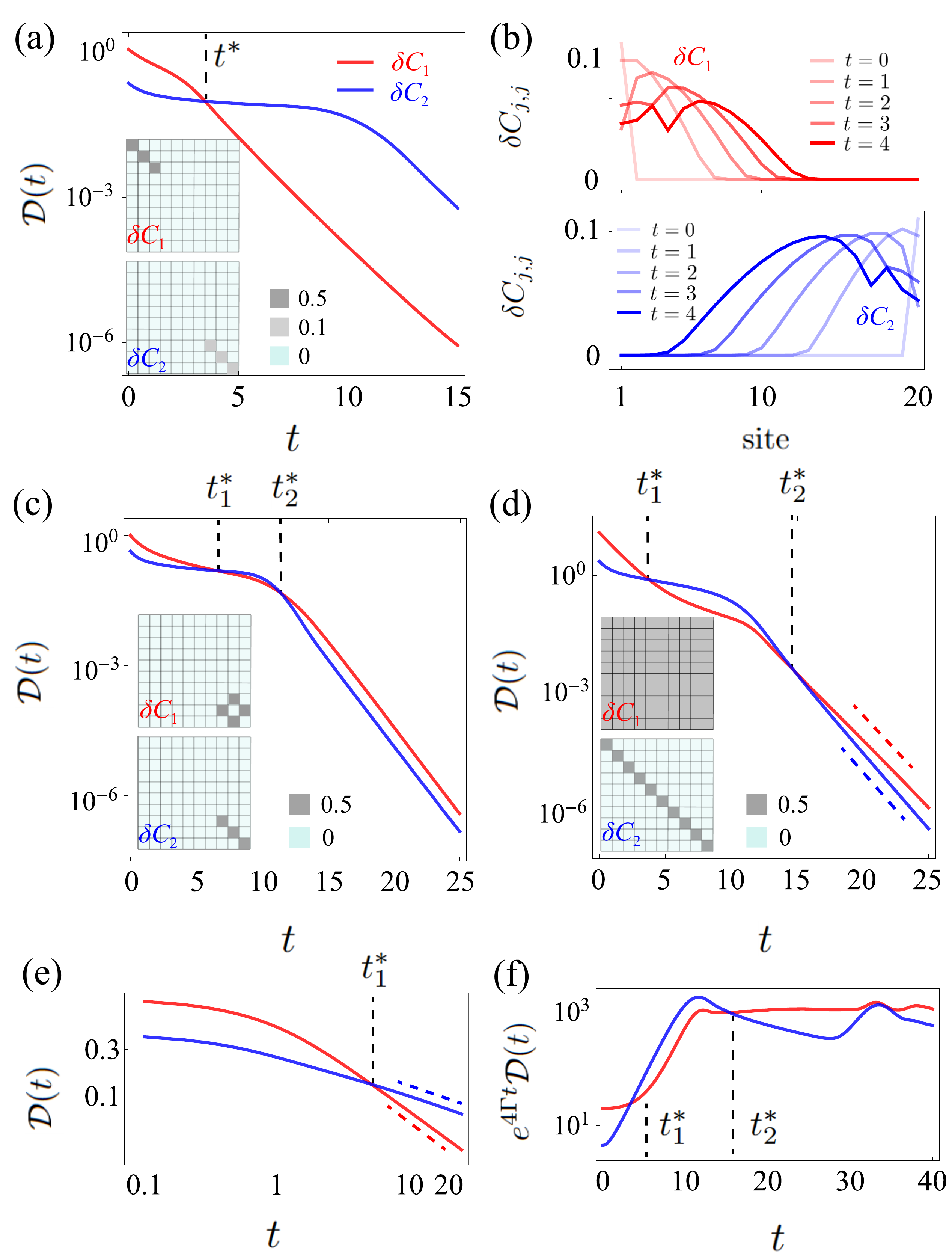}
	\centering
	\caption{Quantum Mpemba effects for diagonal initial states (a-b) and off-diagonal states (c-f).  (b) illustrates the propagation of two initial states in (a). (e) is the same as (c) but calculated in PBC and in Log-Log plot.  The blue and red dashed lines indicate that at early times (around $t_1^*$) the two initial  states relax as ${\cal{D}}(t) \sim t^{-1/2}$ and ${\cal D}(t)\sim t^{-3/4}$, respectively.  (f) is the same as (d) but with ${e^{4\Gamma t}\cal D}$ so that the intrinsic dynamics is revealed. It explains why at long time (around $t_2^*$) the exponential decay rates are different in (d) as indicated by the blue and red dashed lines. }
	\label{me}	
\end{figure} 

{\it Quantum Mpemba effect.--}
Many physical quantities may be used as metrics for measuring the distance between two states \cite{Ares2025quantum}. In this work, we adopt the  Frobenius or Hilbert-Schmidt distance \cite{Dajka2011dis,Kumar2020wish} to quantify the distance from the steady state, which is 
defined as 
\begin{equation}
    \mathcal{D}(t)=\sqrt{{\rm Tr}[\delta C^{\dagger}(t)\delta C(t)]}.
    \label{hs}
\end{equation} 
Mathematically, the Hilbert-Schmidt distance can be understood as the Euclidean distance between two vectorized matrices, and plays a crucial role in diverse problems in quantum information theory, such as the construction of entanglement witnesses \cite{Bertlmann2005optimal,Bertlmann2008geo}, quantum algorithms in machine learning \cite{Vojtech2019exp}, and quantum-state tomography \cite{Sugiyama2013pre,Zhu2011quantum}. It has also been used to observe the strong QME in Markovian systems \cite{Carollo2021exp}
 The QME occurs if there exists a time $t^{*}$ such that the distances for two initial states cross with each other. Calculation of distance based directly on density matrix $\rho$ also gives rise to the same physical results, see SM for details.

We now consider a prototypical model as shown in Fig.~\ref{model} (a) with Hamiltonian given by
\begin{equation}
    H=-J\sum_{j=1}^{L-1} (c_j^\dagger c_{j+1}+c_{j+1}^\dagger c_j).
\end{equation}
Specifically, gain and loss jump operators  take the form 
\begin{equation}
\begin{split}
L_j^g&=\sqrt{\frac{\gamma_g}{2}}(c_j^\dagger +ic_{j+1}^\dagger)\\
L_j^l&=\sqrt{\frac{\gamma_l}{2}}(c_j -ic_{j+1}),
\end{split}
\end{equation}
where $\gamma_{l/g}$ is the dissipation (gain) rate.  These jump operators can be realized within the framework of reservoir engineering \cite{Metelmann2015non,Wanjura2020topological,Porras2019top}.

From Eq.~(\ref{xf}), the effective non-Hermitian Hamiltonian $\mathcal{H}_{{\rm eff}}$ in real space is
\begin{equation}
\begin{split}
    \mathcal{H}_{{\rm eff}}&=-i H_{HN}-\sum_{j=1}^L 2\Gamma c_j^\dagger c_{j}
\end{split}
    \label{xh}
\end{equation}
where $\Gamma=(\gamma_l+\gamma_g)/2$, and $H_{HN}$ is the Hamiltonian for the Hatano-Nelson model \cite{Hatano1996loca,Hatano1997vortex,Hatano1998non}:
\be
   H_{HN}= \sum_{j=1}^{L-1} \left((J+\Gamma)c_j^\dagger c_{j+1}+(J-\Gamma)c_{j+1}^\dagger c_j \right) \label{HHN}
\ee
From Eq.~(\ref{xh}) and (\ref{HHN}), the strength of left hopping is greater than right hopping, thus manifesting a leftward Liouvillian skin effect, as demonstrated in Fig.~\ref{model} (c).
The eigenvalues of effective Hamiltonian $\mathcal{H}_{{\rm eff}}$ under open boundary conditions (OBC) are given by
\begin{equation}
    \lambda_k=-2\Gamma -i \varepsilon_k, 
    \label{eigen}
\end{equation}
with $\varepsilon_k = 2 \Tilde{J} \cos k$ being the eigenvalues of the Hatano-Nelson model. Here, $\tilde{J} = \sqrt{(J-\Gamma)(J+\Gamma)}$,  $k=\pi m/(L+1)$ with $m=0,...,L-1$.  In contrast, for periodic boundary conditions (PBC), the eigenvalues are $\lambda_k^{\rm PBC}= -2\Gamma -2i J \cos k -2\Gamma \sin k$ with $k=2\pi m/L$ with $m=0, 1, \ldots, L$. 
The eigenenergies for OBC and PBC are plotted in Fig.~\ref{model}(b).

%Globally, the stability of quadratic Lindbladians can be quantified with the Liouvillian gap $\Delta={\rm Re}\lambda_k=-2\Gamma$. The Liouvillian gap $\Delta$ is always negative, ensuring that there always exist a steady state. 
%

Assuming  uniform dissipation rate for loss and gain operator and setting $\gamma_l=\gamma_g$, the steady state is simply $C_{ss}=(1/2) \mathbb{I}_{L\times L}$ with $\mathbb{I}$ the $L\times L$ identity matrix \cite{Song2019non}. The evolution of correlation matrix in Eq.~(\ref{dcf}) can be rewritten as:
\begin{equation}
    \delta C(t)=e^{-4\Gamma t}\sum_{k, k'}e^{i(\varepsilon_{k} + \varepsilon_{k'})t}|R_k\rangle\langle L_k|\delta C_0|L_{k'}\rangle\langle R_{k'}|.
    \label{dcf2}
\end{equation}
Note that, under OBC, the real part of the eigenvalues are all the same. Therefore, the argument of QME based on the elimination or reduction of slowest decay decay mode in Eq.~(\ref{rhot}) may not be applicable. 
According to (Eq.~\ref{dcf2}), one would naively expect that all the initial states would decay exponentially  with the same relaxation rate $4\Gamma$. 
However, we will see that, due to the existence of LSE, more complicated relaxation processes would emerge and play a decisive role in QME. 

%We now demonstrate that the LSE provides a natural platform for realizing the QME. In particular, we consider two distinct cases to construct reasonable initial states, built on the LSE, to engineer the QME-II and QME-III.

{\it Case-I.--} First, we consider the simple case in which the initial correlation matrix $C(0)$ is purely diagonal, i.e., without inter-site correlations. 
We choose two different initial states  $\delta C_1$ and $\delta C_2$ as depicted in Fig.~\ref{me} (a) with particles initially localized  near the left and right ends, respectively, and with different distances $\mathcal{D}_1>\mathcal{D}_2$. 
% Obviously, the initial reduced distance $\mathcal{D}_R^1$ is larger than $\mathcal{D}_R^2$ at time $t=0$. In Fig.~\ref{me} (b), for Type-I jump operators, the initial state $C_1$ farther from the steady state relaxes faster than a close one $C_2$, so-called quantum Mpemba effect (QME-II). At time $t<t^*$, the distance $\mathcal{D}_R^1>\mathcal{D}_R^2$, whereas, $\mathcal{D}_R^1<\mathcal{D}_R^2$ for $t>t^*$. In stark contrast for Type-II jump operators where the eigenstates of the effective Hamiltonian $\mathcal{H}_{{\rm eff}}$ are extended, the initial state $C_1$ relaxes faster than $C_2$ at all times with the same decay exponent $2|\Delta|$, as demonstrated in inset of Fig.~\ref{me} (b). 
It is clearly seen that, while $\delta C_1(t)$ experiences simply exponential decay, $\delta C_2(t)$ experiences two distinctive relaxation processes, first algebraic decay, and then exponential decay.  A simple  understanding can be given based on Fig.~\ref{me} (b). 
  For $\delta C_2$, the state propagates from  right to left with the   amplitude accumulating due to the non-Hermitian Hamiltonian $H_{HN}$ with asymmetric hopping and at the same time decays exponentially with uniform rate $4\Gamma$, according to Eq.~(\ref{dcf2}). Altogether, the non-Hermitian amplification compensates the exponential decay, and the total decay becomes instead a algebraic  damping. This algebraic decay has already been discussed in Ref.~\cite{Song2019non}. 
For initial $\delta C_1$,  the state propagates from left to right with diminishing amplitude due to the asymmetric hopping, and  at the same time decays exponentially with rate $4\Gamma$. The total damping rate is therefore larger than $4\Gamma$. Detailed discussions can be found in SM.

%We can also understand the QME illustrated in Fig.~2b from the underlying physical mechanism of the QME-II type discussed in previous works\cite{}. When LSE is introduced, the coefficient term $\langle L^i|\delta C_0|L^j\rangle$ in Eq.~(\ref{dcf}) exhibits a qualitatively distinct behavior for the two initial states $\delta C_1$ and $\delta C_2$.  Explicitly, if the right eigenstates $|R^i\rangle$ are exponentially localized at the left edge, the corresponding left eigenstates $|L^i \rangle$ are localized at the right edge \cite{Yao2018edge}. Therefore, LSE leads to a significant discrepancy in overlaps $\langle L^i|\delta C_0|L^j\rangle$ for the two initial states $C_1$ and $C_2$, resulting in a significant different relaxation at early times, and thus QME-II with  long time evolutions of both initial states sharing the same asymptotic decay rates.

{\it Case-II.--} Next, we  consider a more complex but interesting case when the initial correlation matrix contains off-diagonal matrix elements,  namely, with correlation between different lattice sites. In Fig.~\ref{me} (c-d) we present the results of two different initial $\delta C_1$ with correlations compared with initial $\delta C_2$ with only diagonal elements. Both of (c) and (d) illustrate a new kind of QME characterized by two crossings in the evolution of distance.  
The main feature is that  $\delta C_1$ with correlations usually decays faster than $\delta C_2$ without correlations at early times, but decays relatively slower at longer times.  

At early times, the evolution has not yet propagated to the left edge (c), and therefore the dynamics can be roughly understood from the properties of effective Hamiltonian under PBC. We note first that for translationally invariant state $\delta C(t)$, the distance $\mathcal{D}(t)$ can be calculated in momentum space as $[\mathcal{D}(t)]^2= \sum_{k} |c_k(t)|^2$, where $c_k(t)$ is the Fourier component of $\delta C_{l,m}(t)$. Considering  initial state $\delta C_{l,m}(0)=a_0 \delta_{l,m}$ with only diagonal elements. Its Fourier transform is $c_{k}(0)=a_0$, and its dependence on time reads $c_k(t)=a_0 e^{-4\Gamma t}e^{4\Gamma t \sin k } $ according to Eq.~(\ref{dcf2}). Here we have used the eigenenergies  of effective Hamiltonian (\ref{xh}) under PBC. Converting the summation over $k$ into integration, and keeping only the dominating part near $k=\pi/2$ by letting $k=\pi/2 + \delta k$, one arrives at 
\begin{equation}
    [{\cal D}(t)]^2 \sim  a_0^2 \int e^{-8\Gamma t \delta k^2/2} d(\delta k) \sim a_0^2 t^{-1/2},
\end{equation}
 which leads to algebraic decay  $\mathcal{D}(t)\sim a_0 t^{-1/4}$ for diagonal initial matrix. Next, consider an off-diagonal initial state $\delta C_{l, m}(0)=a_1(\delta_{l, m+1}+\delta_{l+1,m})$. Its Fourier transform is now $c_{k}(0)=2a_1 \cos k $, and its dependence on time is $c_k(t)= 2a_1 e^{-4\Gamma t}e^{4\Gamma t \sin k } \cos k $. %Then  $[\mathcal{D}_{HS}(t)]^2=  (2La_1^2/\pi) e^{-8\Gamma t}\int_{k=0}^{2\pi} e^{8\Gamma t \sin k } \cos^2 k dk$. Again let $k=\pi/2+\delta k$, we have $\int e^{8\Gamma t \sin k } \cos^2 k dk\sim e^{8\Gamma t} \int e^{-8\Gamma t \delta k^2/2 } \delta k^2 d(\delta k)\sim  e^{8\Gamma t} t^{-3/2} $, 
Similar procedure gives $\mathcal{D}(t)\sim \sqrt{2} a_2  t^{-3/4}$. Hence, at early times, both the diagonal and off-diagonal initial state display algebraic decay but with different powers, and the off-diagonal initial state decays faster. In (e), we plot the distances for the same initial states in (c) but under PBC, and find that the decay behaviors are in good agreement with the above argument. Note that under PBC, there will be no second crossing at $t_2^*$, since the second crossing is dominated by the process of state propagating from left edge to right. 

In Fig.~\ref{me}(d), an interesting feature appears that at longer time, the strongly correlated $\delta C_1$ experiences smaller exponential decay rate than the diagonal $\delta C_2$. For better clarity, in (f), we eliminate the  uniform decay by plotting $e^{4\Gamma t} {\cal D}_{1,2}(t)$. For $\delta C_2$, it shows that the dominate mode first propagates  from right to left with accumulating magnitude and then from left to right with diminishing magnitude, with latter introducing an extra decay in additional to the uniform decay rate $4\Gamma$. However, for the strongly correlated state $\delta C_1$, the clear pattern of traveling from right to left and then from left to right is smeared out and therefore, extra decay disappears.

{\it Concluision.--}
In this work, we have proposed a new approach to observe the QME through
 LSE in open quantum systems. We have shown that LSE serves as an ideal platform for realizing the QME and the spatial profile of  LSE provides a straightforward pathway to select the initial state, thereby enabling readily accessible experimental preparation.
 Based on the Lindblad master equation with linear jump operators, the so-called quadratic Lindbladian,  and by monitoring the single particle correlation matrix, we show that the states initially localized on opposite ends display distinctive relaxation process, either algebraic or exponential decay, and the QME can be easily achieved. Our approach circumvents the necessity of fine tuning of initial states or control parameter. The discussion can be directly generalized to the class of Lindblad equation with quadratic jump operators. 
We also illustrate that inclusion of inter-site correlation in the initial state would accelerate the relaxation process compared with states that contain only diagonal elements, and thus constitutes new kind of QME with complex structures. Moreover, the QME discovered here is dominated not by exponential relaxation process but rather by algebraic relaxation process, which confronts the traditional picture that the dominant role is played by the slowest exponential decay mode. Our work unveils the deep connection between QME and LSE, and opens a new avenue of study of QME in thermodynamical system with internal spatial structure.

Note: Upon the completion of the manuscript, we noticed appearance of  Ref.~\cite{Longhi2026}  that discussed the effect of Liouvillian skin effect on quantum Pontus-Mempba effect with a focus on two-step protocol. 

{\it Acknowledgments.--} This Letter was supported by  the National Natural Science Foundation of China (Grant No. 12275075).

\bibliography{ref}

\vspace{0.5cm}

\end{document}